%Paper: hep-ph/9212245
%From: rattazzi%theorm.hepnet@Lbl.Gov
%Date: Wed, 9 Dec 92 17:33:52 PST

\magnification=\magstep1

\font\tenrm=cmr10

\font\nineit=cmti9

\hsize=6.5truein
\vsize=8.5truein

\def\ref{\line{} \bigskip
		\baselineskip=17pt plus 0.5pt
		\noindent {{\bf References}}
		\vskip 0.5 truecm\baselineskip=17pt plus 0.5pt}

\def\gev{\,{\rm GeV}}
\def\mev{\,{\rm MeV}}
\def\tev{\,{\rm TeV}}
\def\li{{\rm Li}}
\def\L{{\cal L}}

\def\cro{{^{\dagger}}}
\def\Mhat{{\hat M}}
\def\ph#1{\hat P_{#1}}                                %widehat

%\headline={\hfil \tenrm\folio \hfil}
\def\hc{{\rm h.c.}}

\def\Tr{{\rm Tr\,}}

\def\eps#1{\epsilon_{#1}}
\def\el{{\langle}}
\def\er{{\rangle}}
\def\Lf{{\Lambda_F}}
\def\Ga#1{{\Gamma_{#1}}}
                                %Cabibbo angle
\def\scab{{\sin \theta_{\rm C}}}

\def\qcab{{\sin^2 \theta_{\rm C}}}
\def\Glr{G_{LR}}
\def\Glrbl{SU(2)_L\times SU(2)_R\times U(1)_{B-L}}     %left right group
\def\bq{{\bar q}}
\def\n+{\eta_+}
\def\n-{\eta_-}

\def\la#1{{\lambda_{#1}}}
\def\ga#1{{\gamma_{#1}}}
\def\gat{{\gamma^\t}}
\def\lat{{\lambda^\t}}

\def\ud{{I_{ud}}}                                          %up-down symm

                        % chiral group

\def\ggd{{G\tilde G}}                                     % anomaly
                                     % M lambda

\def\t{{\rm\scriptscriptstyle T}}
\def\l{{\scriptscriptstyle L}}
\def\r{{\scriptscriptstyle R}}
\def\z{{\scriptscriptstyle Z}}

\def\a{{\scriptscriptstyle A}}
\def\b{{\scriptscriptstyle B}}

\def\f{{\scriptscriptstyle F}}
\def\u{{\scriptscriptstyle U}}
\def\d{{\scriptscriptstyle D}}

\def\n{{\scriptscriptstyle N}}

\def\m@th{\mathsurround=0pt }
\def\leftrightarrowfill{$\m@th \mathord\leftarrow \mkern-6mu
	\cleaders\hbox{$\mkern-2mu \mathord- \mkern-2mu$}\hfill
	\mkern-6mu \mathord\rightarrow$}
\def\overleftrightarrow#1{\vbox{\ialign{##\crcr
	\leftrightarrowfill\crcr\noalign{\kern-1pt\nointerlineskip}
	$\hfil\displaystyle{#1}\hfil$\crcr}}}

\def\ds{\leftrightarrow}

\def\bbr{\overline}

\def\simlt{\mathrel{\lower2.5pt\vbox{\lineskip=0pt\baselineskip=0pt
           \hbox{$<$}\hbox{$\sim$}}}}
\def\simgt{\mathrel{\lower2.5pt\vbox{\lineskip=0pt\baselineskip=0pt
           \hbox{$>$}\hbox{$\sim$}}}}

\font\ssmall=cmr9 scaled\magstep0
\font\bbig=cmbx10 scaled\magstep2
\font\bbbg=cmbx9 scaled\magstep0
\ssmall
%\nopagenumbers
\baselineskip=10truept
\line {\hfill \ssmall LBL-32889}
\line {\hfill \ssmall LMU-13/92}
\vskip 1.0truecm
\centerline{\bbig Inverse Hierarchy Approach to}
\vskip 0.5truecm
\centerline{ \bbig Fermion Masses}
\vskip 1.0truecm
\centerline{{\bbbg Zurab G. Berezhiani} \footnote{$^{1}$}{\ssmall
Alexander von Humboldt fellow.}}
\vskip 0.4truecm
\centerline {\nineit Sektion Physik, Universit\"at M\"unchen, D-8000
M\"unchen 2, Germany}\footnote{}{\ssmall
 E-mail: zurab@hep.physik.uni-muenchen.de, 39967::berezhiani}
\vskip 0.3truecm
\centerline {\nineit and}
\vskip 0.3truecm
\centerline {\nineit Institute of Physics, Georgian Academy of
 Sciences, SU-380077 Tbilisi, Georgia}
\vskip 0.8truecm
\centerline{{\bbbg
Riccardo Rattazzi} \footnote {$^{2}$}{\ssmall INFN fellow. E-mail:
rattazzi\%theorm.hepnet@csa.lbl.gov}}

\vskip 0.4truecm
\centerline{\nineit Theoretical Physics Group,
Lawrence Berkeley Laboratory,}
\vskip 0.3truecm
\centerline {\nineit Berkeley, CA-94720, U.S.A.}
\vskip 1.4truecm
\centerline {\bbbg Abstract}
\vskip 0.3truecm
\midinsert\narrower\baselineskip=12truept
{ The first fermion family might play a special role in
understanding the physics of flavour. This possibility is suggested by
the observation that the up-down splitting within quark families increases
with the family number: $ m_u\sim m_d$, $m_c>m_s$, $m_t\gg m_b$. We construct
a model that realizes  this feature of the spectrum in a natural way. The
inter-family hierarchy is first generated by radiative phenomena in a sector
of heavy isosinglet fermions and then transferred to quarks by means of a
universal seesaw. A crucial role is played by left-right parity and
up-down isotopic symmetry. No family symmetry is introduced.
The model implies  $m_u/m_d>$ 0.5 and the Cabibbo
angle is forced to be $\sim\sqrt{m_d/m_s}$. The top quark is
naturally in the 100 GeV range, but not too heavy: $m_t<$ 150 GeV.

{Inspired} by the mass matrices obtained in the model for quarks,
we suggest an ansatz also including charged leptons.
The differences between $u$-, $d$- and $e$-type fermions are simply
 parametrized by three complex coefficients $\eps{u}$, $\eps{d}$
and $\eps{e}$. Additional consistent predictions  are obtained:
 $m_s$=100-150 MeV and $m_u/m_d<$ 0.75.}

\endinsert
\vskip 1.0truecm
\leftline {\ssmall November 1992}
\vfill
\eject

\baselineskip= 14pt plus 0.5pt
\tenrm

\noindent {\bf 1. Introduction}
\vskip 0.5truecm
One of the major issues in modern particle physics is to understand the
origin of the observed pattern of fermion masses and mixing.
In the Standard Model all the flavour structure is determined by
{\it uncalculable} Yukawa couplings.
A hypothetical Theory of Flavour should allow a calculation of these
couplings, or somehow constrain them. When thinking of such a theory,
one should keep in mind the following qualitative features:
\item{i)}  An inter-family mass hierarchy which is
stronger for the up quarks than for the down quarks [1]. The
plot in Fig. 1 suggests for the $i$-th family quark masses
$$m_i^u \sim\eps{u}^{3-i}m_3^u\quad\quad\quad
 m_i^d \sim \eps{d}^{3-i}m_3^d\eqno(1)$$
where $\eps{u}^{-1}=200-300$ and $\eps{d}^{-1}=20-30$.
\item{ii)} The CKM mixing matrix $V$ is close to the unit matrix,
and it has a hierarchy in its off-diagonal entries:
%$V_{ub}\simeq 10^{-1}V_{cb}$, $V_{cb}\simeq 0.25V_{us}$, $V_{us}\simeq 0.22$.
$V_{ub}\sim \eps{d}^2$, $V_{cb}\sim \eps{d}$ and
$V_{us}\sim \eps{d}^{1/2}$.

The inter-family hierarchy makes the radiative picture of fermion mass
generation [2-4] attractive. In particular, the quark mass pattern (1) can
be interpreted in terms of a {\it charge diagonal} radiative cascade. In
this view $\eps{u}$ and $\eps{d}$ are the loop expansion parameters
respectively in the up and the down sector [4]. Leptons, on the other hand,
seem to evade the simple rule $m_i \sim \eps{}^{3-i}m_3$, and this suggests
that their masses may be more difficult to understand (see Fig. 1).
 We therefore focus first on the quark sector.

When discussing quarks, we can also exploit experimental information on the
CKM matrix. The relation $V\simeq 1$ implies that the Standard Model
Yukawa matrices $\Gamma^{u}$ and $\Gamma^{d}$ are somehow aligned.
This property might hint that some ``isotopic'' symmetry interchanging
$u$- and $d$-quarks  has played a role in the mass generation phenomenon.
(A horizontal family symmetry could also be responsible for the alignment
of $\Gamma$'s. However the idea of an up-down symmetry is simpler.)
Let us suppose that such a symmetry indeed exists, and let us call it
$\ud$. In the low energy theory, i.e. the Standard Model,
$I_{ud}$ plays the role of the usual custodial symmetry [5].
In the limit of exact $I_{ud}$ it would be $\Gamma_u=
\Gamma_d$, so that we would have $V=1$ and the up and down sectors unsplit.
Within the radiative picture the breaking of $I_{ud}$ implies the appearance
of {\it small} mixing angles $V_{ub}\sim \eps{d}^2$ and
$V_{cb}\sim \eps{d}$.\footnote{$^{1)}$}{However, also
$V_{us}$ has to be $O(\eps{d})$. This is a generic problem of the
{\it direct} radiative scenarios [3-4]. As it was shown
in Ref.\ [6], the correct value $V_{us}\approx 0.22$ implies a choice of
parameters which spoils the validity of perturbation theory.}

As far as the spectrum is concerned, Fig.\ 1 shows that
$I_{ud}$ is more badly broken for the heavier families. By focusing on the
first family, we don't see a big violation of $I_{ud}$: $m_u/m_d=O(1)$.
What we consider interesting is the case in which
$\ud$ becomes exact at very high energies, namely the case of spontaneous
or soft breakdown. With this picture in mind, it is suggestive to think that
the masses of $u$ and $d$ are somehow related to an energy scale $\Lambda_1$
at which $I_{ud}$ is still  good, while the masses of the second and third
families are respectively related to lower scales $\Lambda_2$, $\Lambda_3$
at which $I_{ud}$ is no longer as good. Thus, the first family seems to play
a special role in mass generation. This is directly opposite to
the common  radiative picture [2-4], in which the third family (i.e. the
heaviest) is the starting point. However, by taking the top and bottom
masses as the {\it seeds} of mass generation, it is difficult to
understand their large splitting.\footnote {$^{2)}$}{In Ref.\ [3-4]
the large $t-b$ splitting is introduced at tree level by hand. A difficulty
of {\it charge diagonal} radiative models is also the appearance of
flavour-changing neutral currents (FCNC), generally contradicting the
experimental bounds [4].} On the other  hand, if we could start from $u$
and $d$, we would have better chance to understand $t$-$b$ splitting. Suppose
that $u$ and $d$ are indeed the starting point, and that an expression like
eq.\ (1) holds for $1/m_i$ instead of $m_i$, namely
$$1/m_i^{u,d}\sim \eps{u,d}^{i-1}/m.\eqno(2)$$
Then we have $m_u\sim m_d\sim m$ and
$m_t/m_b\sim (\eps{d}/\eps{u})^2\sim (m_c/m_s)^2\gg 1$. In this way, the
splitting between up and down quark masses in Fig.\ 1 is understood by means
of one parameter $(\eps{d}/\eps{u})>1$. We call the above
formula for $1/m_i$ the {\it inverse hierarchy pattern}.

In this paper we explore the idea of inverse hierarchy for fermion masses.
In Section 2, we discuss an illustrative model for the quark sector.
The model is a particular case of the class of models discussed in Ref.\ [7].
These models naturally avoid the key problems of the
previous models of radiative mass generation (see footnotes$^{1,2)}$).
A set of isosinglet heavy quarks, $Q$-fermions, in a one to one correspondence
with the ordinary ones ($q$'s), is introduced. The mass matrices of the $q$'s
$\hat{M}_u$ and $\hat{M}_d$ are induced by a seesaw mixing with the $Q$'s [8].
In terms of the mass matrices $\Mhat_{\u,\d}$ of the $Q$'s we have essentially
the inverse proportionality\footnote{$^{3)}$}{The seesaw mechanism [9],
universally extended to quarks and charged leptons was
also explored in the papers in Ref.\ [10,11]. The inverse hierarchy,
however, corresponds to the spirit of the original paper [8], where
it was first suggested,  though in the context of horizontal symmetry.}
 $\Mhat_u\propto \Mhat_\u^{-1},~~\Mhat_d\propto \Mhat_\d^{-1}.$
%$$\Mhat_u\propto \Mhat_\u^{-1}\quad \quad
% \Mhat_d\propto \Mhat_\d^{-1}.\eqno(2)$$
Thus the smallest masses $m_u$ and $m_d$ are indeed related to the highest
scale: $m_u^{-1}\propto M_\u$ and $m_d^{-1}\propto M_\d$, where $M_\u$ and
$M_\d$ are the largest eigenvalues respectively in $\Mhat_\u$ and $\Mhat_\d$.
These matrices are generated  by a radiative mechanism analogous to that in
Ref.\ [4]: the largest eigenvalues $M_{\u,\d}$ appear at tree level, while
the other ones are induced by charge diagonal loop effects. Moreover, the
relation $M_\u\sim M_\d$ arises naturally as a consequence of an up-down
symmetry. Then  $m_u\sim m_d$, so that the pattern of eq.\ (2) is realized.

In the model we discuss below, the field content and the $I_{ud}$ symmetry
imply strictly $M_\u=M_\d$. The model also has $P_{LR}$ and $CP$ symmetries.
These, together with $I_{ud}$, are softly (or spontaneously) broken in the
scalar potential. The effects of this breaking are reflected in the fermion
sector  via loop corrections {\it only}. As a consequence, the mass matrices
are more constrained than those in Ref.\ [7]. In particular, the splitting
between $u$ and $d$ is related to the enhancement of the Cabibbo
angle above its natural radiative value $O(\epsilon_d)$.
%$\scab^{\rm rad}\sim m_d/m_s$.
The consequences, including formulae for
the top quark mass and the Cabibbo angle, are discussed in Section 3.

The quark mass matrices obtained in the model are a very attractive
realization of the inverse hierarchy. Inspired by the form of those matrices,
in Section 4 we simply postulate an inverse hierarchy ansatz, which includes
charged leptons. It turns out that the first family $(u,d,e)$ really plays
the role of a {\it mass unification point}, with its splittings understood
by the same mechanism that enhances $V_{us}$.
This ansatz provides remarkable predictions for the light quark masses,
consistent with the present knowledge on these quantities.

Finally, in Section 5 we discuss our results.

\vskip 1.0truecm

\noindent {\bf 2. A model}
\vskip 0.5truecm
Let us consider the simple left-right symmetric model, based on the gauge
group $\Glr=[SU(2)_L\times U(1)_L]\times
[SU(2)_R\times U(1)_R]\times U(1)_{\bbr B-\bbr L}$,
suggested in [7]. The left- and right-handed components of usual quarks $q_i=
(u_i,\,d_i)$ and their heavy partners $Q_i=U_i,\,D_i$ are taken in the
following representations
$${\eqalign {q_{\l i}&\sim (1/2,0,0,0,1/3)\cr
U_{\l i}&\sim(0,1,0,0,1/3)\cr D_{\l i}&\sim(0,-1,0,0,1/3)\cr}}
\quad\quad\quad\, {\eqalign {q_{\r i}&\sim(0,0,1/2,0,1/3)\cr
 U_{\r i}&\sim(0,0,0,1,1/3)\cr
D_{ \r i}&\sim(0,0,0,-1,1/3)\cr}} \eqno(3)$$
where $i=1,2,3$ is the family index (the indices of the colour $SU(3)_c$ are
omitted). We also introduce an additional family of quarks
\footnote{$^{4)}$}{ Notice that, by adding the obvious lepton fields
(with $\bbr B-\bbr L=-1$) to the quarks of eqs.\ (3-4),
the theory is free of gauge anomalies.}
$$\eqalign {p_\l&\sim(0,-1/2,0,3/2,1/3)\quad\quad\quad \,p_\r\sim
(0,3/2,0,-1/2,1/3)
\cr n_\l&\sim(0,1/2,0,-3/2,1/3)\quad\quad\quad\,n_\r\sim(0,-3/2,0,1/2,1/3).\cr}
\eqno(4)$$
 The scalar sector of the theory is given by
$${\eqalign {H_\l&\sim(1/2,0,0,1,0)\cr
T_{u\l}&\sim(0,-2,0,0,-2/3)\cr T_{d\l}&\sim(0,2,0,0,-2/3)\cr\phi&\sim
(0,1/2,0,-1/2,0)\cr\Omega_u&\sim(0,1/2,0,1/2,-1)\cr}}
\quad\quad\quad\, {\eqalign {H_\r&\sim(0,1,1/2,0,0)\cr T_{u\r}&
\sim(0,0,0,-2,-2/3)\cr T_{d\r}&\sim(0,0,0,2,-2/3)\cr\Phi&\sim(0,2,0,-2,0)\cr
\Omega_d&\sim(0,-1/2,0,-1/2,-1)\cr}}\eqno(5)$$
 where the $T$-scalars are colour triplets or sextets. We impose
 $CP$, $P_{LR}$ and  $\ud$ discrete symmetries. The left-right symmetry
 $P_{LR}$ [12], which is essentially parity, and $CP$ act in the usual way.
The ``up-down'' symmetry $\ud$ is defined by
$$\eqalign {U_{\l,\r}&\ds D_{\l,\r}, \qquad  p_{\l,\r}\ds n_{\l,\r},
\qquad A_{\l,\r}^\mu \rightarrow -A_{\l,\r}^\mu,\cr
H_{\l,\r}&\ds \tilde H_{\l,\r},\qquad T_{\l,\r}^u\ds T_{\l,\r}^d,\qquad \phi\ds
\phi^*,\qquad \Phi\ds \Phi^*,\qquad\Omega_u\ds\Omega_d\cr}
\eqno(6)$$
where $A_{\l,\r}^{\mu}$ are the gauge bosons of $U(1)_{L,R}$. Then the most
general Yukawa interactions, consistent with gauge invariance,
$\ud$, $P_{LR}$ and  $CP$ are
 $$\eqalign {\L_1&=\Gamma^{ij}\bigl \{\bq_{\l i}U_{\r j}\tilde H_\l+\bq_{\l i}
D_{\r j}H_\l +(L\ds R)\bigr \}+\hc\cr
\L_2&={\la{ij}\over 2}\bigl \{T_{u\l}U_{\l i}C U_{\l j} +
T_{d\l}D_{\l i}C D_{\l j}+(L\ds R)\bigr \} +\hc\cr
\L_3&=\bigl \{h_i(\phi^*\bar U_{\l i}p_\r +\phi\bar D_{\l i}n_\r)+
h(\Phi^*\bar p_\l p_\r+\Phi \bar n_\l n_\r)+(L\ds R)
\bigr \}+\hc\cr}
\eqno(7)$$
where $C$ is the charge conjugation matrix. The coupling constants $h$, $h_i$,
$\lambda_{ij}$, $\Gamma_{ij}$ $(i,j=1,2,3)$ are real, due to $CP$-invariance.
In what follows we do not make any particular assumption on their flavour
structure. We only suppose that they are typically $O(1)$, just like the
gauge coupling constants. Notice that $\lambda$ is antisymmetric (symmetric)
%$\lambda_{ij}= -\lambda_{ij}$,
when the $T$-scalars are colour triplets (sextets). Without loss of
generality,  by a suitable redefinition of the fermion basis, we
can always take
$$h_{2,3}=0,~~~~~\lambda_{13}=0,~~~~~\Gamma_{ij}=0 {\rm ~~ if~~ } i<j~~.
\eqno(8)$$
We assume that the discrete symmetries $CP$, $P_{LR}$ and  $\ud$ are
{\it softly} broken by bilinear and trilinear terms in the Higgs potential
\footnote{$^{5)}$}{Actually, these symmetries can be broken spontaneously by
suitable scalars [7]. }. The trilinear terms are given by
$$ \L_4=
\Lambda_u T_{u\l}^* T_{u\r}\Phi+\Lambda_d T_{d\l}^*T_{d\r}\Phi^* +\hc
\eqno(9)$$
where $\Lambda_{u,d}$ are mass dimensional coupling constants. We suppose that
these couplings are complex, and also that $\Lambda_u\not = \Lambda_d$.
In this way, $CP$, $P_{LR}$ and $I_{ud}$ are broken.

The v.e.v.'s $\el\Phi\er=v_{\Phi}$, $\langle\phi\rangle=
v_{\phi}$ and $\el\Omega_u\er$ break the
$U(1)_L\times  U(1)_R\times U(1)_{\bbr{B}-\bbr{L}}$ symmetry down to
the usual $U(1)_{B-L}$: $B-L=Y_L+Y_R+\bbr{B}-\bbr{L}$.
(Notice that we must have $\el\Omega_d\er =0$.) Then, interactions $\L_3$
in eq.\ (7) give rise to mass terms for $p$ and $n$, and also
for the $Q$-fermions of the ``first family'' $U_1$ and $D_1$. It
is interesting to consider the limiting case $v_\Phi \gg v_\phi$.
In this case, the tree-level mass eigenvalues are given by the seesaw
formula: $M_p=M_n\simeq hv_{\Phi}$ and $M_{\u_1,\d_1}=M\simeq h_1^2
v_{\phi}^2/hv_{\Phi}$. Because of the relation $M\ll M_{p,n}$,
the mass matrices obtained after including radiative
effects depend, in general, on a smaller number of parameters.
In fact, we shall see below that a {\it soft} hierarchy
$hv_\Phi/h_1v_\phi\sim 3$ is already enough for our purposes.

The coloured scalars $T_{u\l}$,$T_{u\r}$ and $T_{dL}$,$T_{dR}$ get mixed due
to interaction  terms in eq.\ (9). At this point the interactions in $\L_2$
trigger the radiative mass generation chains  $U_1\to U_2\to U_3$
and $D_1\to D_2 \to D_3$ (see Fig.\ 2). The two-loop corrected mass
matrices for the $Q$-fermions can be written as
$$\eqalign {\Mhat_{\u,\d}=&M\Bigl \{\ph{1}+\eps{u,d}\,\,\la{}^\t\,
\ph{1} {\lambda}+
C_{u,d}\vert\eps{u,d}\vert^2\;\lat^2 \,\ph{1}\,{\lambda}^2\cr
&+\eps{u,d}\la{}^\t\bigl (\beta_{u,d}^\l\lat^2+\alpha_{u,d}^\l\Ga{}^\t\Ga{}
\bigr )\ph{1}\la{} +\eps{u,d}\la{}^\t\ph{1}
\bigl (\beta_{u,d}^\r\la{}^2+\alpha_{u,d}^\r\Ga{}^\t\Ga{}\bigr )\la{}
\Bigr \}\cr}\eqno(10)$$
  where $\ph{1}={\rm diag}(1,0,0)$ is the tree-level term. In terms of the
mixing angle  $\xi_{u,d}$ and mass eigenvalues $M_{+}^{u,d}, M_{-}^{u,d}$
of the scalars $T_\l^{u,d}$-$T_\r^{u,d}$, the expansion parameters are
$$\eqalign {\eps{u,d}&=C_T{e^{i\omega_{u,d}}\over 16\pi^2}
\sin (2\xi_{u,d}) \ln r_{u,d}, \qquad~~~r_{u,d}=(M_+^{u,d}/M_-^{u,d})^2,\cr
C_{u,d}&={1\over 2}\bigl (1+H(r_{u,d})\bigr )~.\cr}\eqno(11)$$
Here $\omega_{u,d}$ are the arguments of the mixing terms $\Lambda_{u,d}$
in eq.\ (9), while $H$ is a real function given in the Appendix.  $C_T$ is a
colour factor equal to $1/2$ for a triplet $T$ and to $1$ for a sextet $T$.
The first three addenda in eq.\ (10) represent respectively the contributions
of graphs (a), (b) and (c) of Fig. 2. They are rank 1 matrices
determining the eigenvalues of $\Mhat$ to be $O(M)$, $O(\eps{} M)$ and
$O(\eps{}^2 M)$. The terms proportional to $\beta_{u,d}$ and $\alpha_{u,d}$
in eq.\ (10) arise from the graphs in Figs.\ 2d-e.
Apart from pieces which are formally 3-loop,
i.e. $O(\eps{}\alpha^2,\eps{}\beta^2,\eps{}\alpha\beta)$, these terms simply
determine a multiplicative redefinition of the 1-loop contribution (Fig. 2b).
(We remind that $\alpha$'s and $\beta$'s are loop coefficients $\sim
1/16\pi^2$.)
This means that in 2-loop order their effect on the eigenvalues is only
multiplicative, i.e. $M_i\to M_i(1+O(\alpha,\beta))$ [3].
For our purposes, these terms are therefore negligible.

Some comments
on eqs.\ (10-11) are in order. Eq.\ (11) is valid in the regime $M^2\ll
{M_{+}^{u,d}}^2, {M_{-}^{u,d}}^2\ll M_p^2$. In this regime the loop-diagrams
of Fig. 2b-c are dominated by loop momenta between $M_{-}^{u,d}$ and
$M_{+}^{u,d}$. Thus, the result is completely determined by the ratios
$r_{u,d}$, and the predictivity is increased. The choice of the intermediate
range for $M_\pm^{u,d}$ is also motivated by the fact that both for
$M_\pm^{u,d}<M$ and for $M_\pm^{u,d}>M_p$ the loop integrals are suppressed.
In the first case the suppression factor is given by $\sim (M_+^{u,d}/M)^2$,
in the second by $\sim (M_p/M_-^{u,d})^2$. (In fact $M_p$ acts as a cut-off
of the mass insertions in Fig. 2 [4].) Notice also that $(M/M_p)^2\simeq
(h_1v_\phi/hv_\Phi)^4\ll 1$ is already implied by our assumption
$v_\Phi>v_\phi$. For instance, $h_1v_\Phi/hv_\Phi\sim 3$ already gives
$(M/M_p)^2 \sim 10^{-2}$.

The function $H(r)=H(1/r)$ has the maximum at $r=1$ and goes to zero
monotonically when $r\to \infty$ (see Appendix). However, in a reasonable
range $(1<r<10)$ it is $H(r) \simeq 0.3$ with good accuracy. Thus, in what
follows we shall set $C_{u,d}=C\simeq 0.65$. Notice that the two loop term
proportional to $C_{u,d}$ in eq.\ (10) is real. This is because the mass
insertions
on the two scalar lines in Fig. 2c are complex conjugate of each other.

As a final comment, notice that, in the basis of eq.\ (8), we have
$$\Mhat \propto \left (\matrix {O(1)&O(\eps{})&O(\eps{}^2)\cr O(\eps{})&
O(\eps{})&O(\eps{}^2)\cr O(\eps{}^2)&O(\eps{}^2)&O(\eps{}^2)\cr}\right )
\eqno (12)$$
Then, in lowest order, the eigenvalues are given by the diagonal entries of
$\Mhat $, and their ratios are $O(1):O(\eps{}):O(\eps{}^2)$.

\vskip 1.0truecm

\noindent {\bf 3. The quark mass matrices}
\vskip 0.5 truecm
We now discuss the masses of the ordinary quarks $q$.
The v.e.v's $\el H_\l\er=(0,v_\l)$ and $\el H_\l\er=(0,v_\r)$, with
$v_\r\gg v_\l=174$ GeV, break the intermediate $\Glrbl$ symmetry down to
$U(1)_{\rm em}$, and at the same time mix the
$q$'s to the $Q$'s via the interactions $\L_1$ in eq.\ (7).
The mass matrix of the $u$-type quarks is written in block form as
 $$(\bar u,\bar U)_\l\left (\matrix {0 &v_\l
 \Gamma\cr v_\r\Ga{}^\t &\Mhat_\u\cr}
\right )\left (\matrix {u\cr U\cr}\right )_\r \eqno(13)$$
 and analogously for the $d$-type ones. When
 $\Mhat_{\u,\d}\gg v_\r$, the resulting mass matrix  $\Mhat_{u,d}$
for the ordinary quarks is given by the seesaw formula [8-10]
$$\Mhat_{u,d}=v_\l
v_\r \Gamma \Mhat_{\u,\d}^{-1}\,\Ga{}^\t \eqno(14a)$$
In this way the inverse proportionality of eq.\ (2) is realized [7,8].
The seesaw limit $M_{\u,\d}\gg v_\r$ is certainly very good for light quarks,
since their masses must be much smaller than $v_\l$. However, since
$m_t=O(v_\l)$, we expect the mass of its Q-partner
$M_\t$ to be of the order of $v_\r$ (remember that the $\Gamma^{ij}$ are
considered to be $O(1)$). Thus, to evaluate $m_t$, we need the mass matrix
without the  restriction $\Mhat_\u\gg v_\r$. This is given by
$$(\Mhat_{u,d}\cro \Mhat_{u,d})^{-1}= v_\l^{-2}\Gamma^{-1}\bigl \{
v_\r^{-2}\Mhat_{\u,\d}\cro (\Gamma\Ga{}^\t)^{-1}\Mhat_{\u,\d}~+~1\bigr \}
( \Ga{}^\t)^{-1}.\eqno(14b)$$
Notice that, when $v_\r\gg\Mhat_{\u,\d}$, this equation gives the obvious
result $\Mhat_{u,d}=v_\l\Gamma$. On the other hand, when
$v_\r\ll \Mhat_{\u,\d}$, eq.\ (14b) reduces to the seesaw formula.

The v.e.v. $v_\r\sim M_\t$ plays the role of the Flavour scale $\Lf$.
Below the scale $v_\r$ the effective theory is the minimal standard model.
In fact $v_\r \Gamma \Mhat_{\u,\d}^{-1}\,\Ga{}^\t=\Mhat_{u,d}/v_\l$ are
the standard model Yukawa matrices $\Gamma_{u,d}$ arising at the decoupling
of the heavy $Q$ states.
The $Q$-fermion masses are such that $M_\t \ll M_\b \ll ...\ll M_\u=M_\d=M$.
Then the ratio $M/v_\r$ is given by $v_\l/m_u\sim 10^{5}$.
Obviously, FCNC's are well controlled in such a theory
[8] due to natural flavour conservation within the standard model [13].
Flavour changing effects are suppressed by inverse powers of the $Q$ masses.
Phenomenologically, the most stringent constraints come from $K^0-\bar{K}^0$
mixing and $K_\l\rightarrow \mu^{+}\mu^{-}$. It turns out that
the typical scale of suppression for these processes is given by $M$.
This is readily seen in basis (8): the light quarks
$u,d$ mix only with the heaviest $Q$-fermions $U,D$.
Consistency with the data is obtained for $M>100\tev$.
Since $M/v_\r\sim 10^5$, we have that $v_\r\sim 1-10$ TeV is already
consistent with the experimental data. Therefore, it is in
principle possible to have the $W_\r$ bosons and the lightest
$Q$-fermion, $T$, in the TeV range, with implications at SSC/LHC.

It is useful to first study $\Mhat_{u,d}$ in the seesaw limit (14a).
The exact formula (14b) will only be relevant to evaluate $m_t$. Once again,
the inverse matrices are easier to analyse.  Neglecting $\alpha$
and $\beta$ terms, as explained above, eq.\ (10) gives
$$ (\Mhat_{u,d})^{-1}=m^{-1}\Bigl \{\ph{1}+\eps{u,d}\,\, \gamma^\t\,
\ph{1}\,\gamma+C\vert\eps{u,d}\vert^2\,\, \gat^2\, \ph{1}\,\gamma^2
+\cdots \Bigr \}\eqno(15)$$
where $\gamma=\Ga{}\lambda\Ga{}^{-1}$ and $m=\Gamma_{11}^2 v_\l v_\r/M$.
We also used the relation $(\Ga{}^\t)^{-1}\ph{1}\Gamma^{-1}
=\Gamma_{11}^{-2}\ph{1}$, true in the basis of eq.\ (8). Notice that,
in general, $\ga{}$ is no longer symmetric or antisymmetric. However
$\Tr \ga{}=\Tr \la{}$, Det$\ga{}$=Det$\la{}$ and, because of eq.\ (8),
$\ga{13}=\la{13}=0$. Eq.\ (15) is then written in the same form as eq.\ (12),
which is easily diagonalized to lowest order in $\eps{}$.

 There is, however, a subtlety involving the Cabibbo angle. Let us consider
eq.\ (15) restricted to the first two families at $O(\eps{})$. We have
$$(\Mhat )^{-1}=m^{-1}\left \{\left (\matrix {1&0\cr 0&0\cr}\right )+
\eps{}\left (\matrix {\ga{11}^2&\ga{11}\ga{12}\cr\ga{11}\ga{12}&\ga{12}^2
}\right )\right \}.\eqno(16)$$
This gives, in lowest order, $\scab\simeq
\eps{d}\ga{11}\ga{12}$ and $d/s\simeq \eps{d}\ga{12}^2$. These relations
imply $\eps{d}\ga{11}^2\simeq \qcab (s/d)\simeq 1$, which looks like  trouble
for perturbation theory [6-7]. However, $\ga{11}=\la{12}\Ga{21}/\Ga{22}$
is not a coupling constant: its value can be enhanced by the ratio
$\Ga{21}/\Ga{22}$, without trouble for perturbativity [7]. This is an
advantage of the seesaw mechanism: the ``sandwiching'' between $\Gamma$'s
in eq.\ (14a)
%of $\lambda$, $\lambda\rightarrow \gamma=\Gamma\lambda\Gamma^{-1}$,
%the relative magnitude of the entries in $\Ga{}$ can
allows the mass matrix in eq.\ (15) to deviate
a little bit from the {\it rigid}
radiative form of eq.\ (12). In particular, when $\Ga{21}$ is larger than
$\Ga{11}$ and $\Ga{22}$, the 1-2 entry gets enhanced. This effect is
reflected in eq.\ (15-16) by the fact that $\ga{11}$ is {\it large}.
Notice that, by supposing $\eps{d}\la{12}^2\sim d/s=1/20$, we only need
$\Ga{21}/\Ga{22}=4-5$.

To accommodate a large Cabibbo angle, the matrix (15) must be diagonalized
considering that $\eps{d}\ga{11}^2=O(1)$. With an obvious notation, we indicate
by $u,d,...$ the quark mass eigenvalues at the flavour scale $\Lambda_F$.
Then we get
$$\eqalign{ (m/u)&=|1+\eps{u}\ga{11}^2| \cr (m/d)&=|1+\eps{d}\ga{11}^2|\cr}
\qquad \eqalign {(m/c)&={{|\eps{u}|\ga{12}^2}\over {|1+\eps{u}\ga{11}^2|}}
 \cr (m/s)&={{|\eps{d}|\ga{12}^2}\over{ |1+\eps{d}\ga{11}^2|}}\cr}
\qquad \eqalign {(m/t)&=C|\eps{u}|^2\ga{12}^2\ga{23}^2\cr
(m/b)&=C|\eps{d}|^2\ga{12}^2\ga{23}^2\cr}\eqno(17a)$$
and the Cabibbo angle is
$$\eqalign {V_{us}&=\ga{11}\ga{12}\left |\bigl ({d\over m}\bigr )
\eps{d}e^{i\delta_d}-\bigl ({u\over m}\bigr )\eps{u}
e^{i\delta_u}\right |=\ga{11}{\sqrt {|\eps{d}|}}{\sqrt {d\over s}}
\,\left |1-{s\over c}e^{i(\omega_u+\delta_u-\omega_d-\delta_d)}\right |\cr
\delta_{u,d}& ={\rm arg }(1+\eps{u,d}^*\ga{11}^2).\cr}\eqno(17b)$$
By dropping terms proportional to $\eps{u,d}\ga{11}^2$ in the above
equations, we would get the naive lowest order result. (Notice that we have
kept also $\eps{u}\ga{11}^2$ terms. However, we shall see that
$\eps{u}\ll\eps{d}$, so that those terms are not relevant). Notice also that
the third family masses are simply given by the 3-3 entry in eq.\ (15), i.e.
they do not get relevant contributions from off-diagonal terms. This is
because their mixings $V_{cb}$ and $V_{ub}$ must be of the natural
{\it radiative} size, namely $V_{cb}\sim \eps{d}\sim d/s,s/b$ and
$V_{ub}\sim \eps{d}^2\sim d/b$.

The following comment is in order. In contrast to the quantities in eqs.
(17), $V_{cb}$ and $V_{ub}$ get lowest order contributions which are
not expressed in terms of $\eps{}$. One of these comes from
Fig. 2d-e: these graphs contribute in lowest order to $V_{cb}$ and
$V_{ub}$, but not to the quantities in eq.\ (17).
In particular,
$V_{cb}$ gets contributions $O(\alpha, \beta)$, whereas $V_{ub}$ gets
$O(\eps{}\alpha, \eps{}\beta)$. This is consistent with the radiative picture
of mixings, since $\alpha,\beta$ are loop parameters $\sim\eps{}$.
%$V_{cb}\sim$ 1-loop and $V_{ub}\sim$ 2-loop.

Another contribution comes from the renormalization of $\Gamma$'s.
These effects are given at 1-loop by triangle graphs involving three
Yukawa vertices in $\L_1$ in eq. (7). The mass splittings between
up and down $Q$-fermions determine this renormalization to differ by a finite
amount in the two sectors.
This is an additional source of mixing. It is remarkable, however, that
a contribution appears at 1-loop {\it only} in the mixing
between the second and third family, namely in $V_{cb}$.
This is a consequence of the fact that the first family ($u,d$) is only
connected to the first heavy family ($U,D$), which is unsplit in lowest
order. Then the vertex graphs involving $u$ and $d$ are equal and do not
contribute to the mixing in lowest order: the relation $M_\u=M_\d$
protects $V_{us}$ and $V_{ub}$ from 1-loop corrections and thereby
maintains $V_{ub}$ as a two loop effect.

The above observation selects the present model from the wide class
discussed in Ref. [7], not only as a predictive one (see below), but also
as the {\it only} one in which the mixing angles $V_{cb}$ and $V_{ub}$
follow the natural {\it radiative} pattern: $V_{cb}\sim$ 1-loop,
$V_{ub}\sim$ 2-loop.\footnote{$^{6)}$} {The relation $M_\u=M_\d$ does
not generally hold in the models of Ref.\ [7]. It was considered to
be broken at tree level. Consequently, in order to have a small enough
$V_{ub}$, some Yukawa coupling in $\L_1$ (namely $\Gamma_{31}$) must be
taken about one order of magnitude smaller then the others.
In Ref.\ [7] the role of the renormalization of $\Gamma$'s had
been overlooked.}

Let us now discuss the implications of eqs.\ (17).
We have seen that $V_{us}$ is enhanced by seesaw effects above
the {\it radiative} value $O(\eps{d})$.
Eqs.\ (17) show the connection between the {\it large} value of the Cabibbo
angle and the $u$-$d$ splitting. Neglecting terms of order
$|\eps{u}/\eps{d}|$, we get
$$V_{us}\simeq {\sqrt {{d\over s}\left |1-{u\over d}e^{i\delta_d}\right |}}
\eqno (18)$$
Thus, without any hypothesis on $\delta_d$, we have $V_{us}$
in the right range: $|V_{us}|\sim \eps{d}^{1/2}$. For instance, for $u/d\leq
0.6$ and $s/d=20$ [1], we obtain $0.14\leq |V_{us}|\leq 0.26$. Notice
that the experimental value of $|V_{us}|$ requires large values $\delta_d
\sim 1$. On the other hand, $\delta_d$ contributes
appreciably to the $CP$ phase in the CKM matrix, even though there are other
independent contributions. These are related to the effects on $V_{ub}$ and
$V_{cb}$ that we discussed above. However, barring
accidental cancellations, we can conclude that the model favours a large
$CP$-phase, in accord with experiments [14].

Let us now discuss the mass eigenvalues. From the masses of the first two
families in eq.\ (17a) we get
$$\bigl |{\eps{u}\over \eps{d}}\bigr |={{sd}\over {cu}}.\eqno(19)$$
On the other hand $b/t=|\eps{u}/\eps{d}|^2$, so that we have a mass
formula
$${t\over b}=\bigl ({{uc}\over {ds}}\bigr )^2.\eqno(20)$$
This equation, valid in the seesaw limit of eq.\ (15), is an appealing
expression for the top mass. However, even when exact, eq.\ (20)
does not translate into a sharp top mass prediction, because of our
poor experimental knowledge of $R=(uc/ds)^2$. By using the correct
mass matrix of eq.\ (14), eq.\ (20) is reduced to an inequality
$$\bigl (1/\la{t}\bigr )^2=\bigl (1/\la{b}R\bigr )^2+\bigl
(1/\Ga{33}\bigr )^2\geq\bigl (1/\la{b}R\bigr )^2\eqno(21)$$
where $\la{t,b}=m_{t,b}/v_\l$ are the standard model Yukawa couplings.
Eq.\ (21) is valid at the ``Flavour scale'' $v_\r$, and, to discuss its
implications, the running of masses needs to be considered. For our
purposes, we only need to consider the combined effects of QCD and of the top
Yukawa coupling [15] (electroweak gauge couplings modify our conclusions
by less than 5$\%$). In what follows we assume $\alpha_s(M_\z)=0.11$ and
use the estimate of the QCD running of masses given in ref. [16] at 2-loop
accuracy. There are two implications from eq.\ (21).

 {i)} {\underbar {An upper bound on the top mass}}. This value is determined
by the maximal allowed values for $R$ and $\Ga{33}$. Taking
$m_u/m_d\leq 0.8$ and $m_c/m_s\leq 10.7$ from the values given in Ref. [1,17]
we have $R<R_{\rm max}=74$ (Notice that $R$ is not affected by the running of
masses, when electroweak couplings are neglected [15].) On the other hand,
$\Ga{33}$ is bounded by requiring the perturbativity of the model:
$\Ga{33}<0.5-1$. This upper bound requires some explanation. From the
definition of $\ga{}$, and from the basis choice eq.\ (8), we
have\footnote{$^{7)}$}{ Actually this relation is true only in the case of a
triplet $T$. However, the consequences in the case of a sextet do not differ
appreciably.} $\Ga{21}/\Ga{33}=(\eps{d}\ga{11}\ga{23})/(\eps{d}
\la{12}\la{23})$. Then, from the expressions for $d/s$ and $\scab$ in
eq.\ (17), we have $\eps{d}\ga{11}^2\simeq 1$, while from the b-mass we have
$\eps{d}\ga{23}^2\sim 0.06$. On the other hand, $\eps{d}\la{12}\la{23}$ is a
true loop expansion parameter in eq.\ (10), so that we demand that it be
smaller than $0.1-0.05\sim d/s$. Putting all these constraints together we
get $\Ga{21}/\Ga{33}>2-4$. Finally, by demanding the reasonable perturbative
upper bound $\Ga{21}<2$ we obtain the bound  $\Ga{33}\simlt 0.5-1.0$.

 The results are shown in Table 1a for different values of  $\Lf$.
Notice that for $\Lf=10^{12} \gev$ the heaviest states of
the model lay just below the Planck mass. However, for the sake of
demonstration we also kept $\Lf=10^{16} \gev$.
As we see the top quark cannot reasonably be heavier
than 150 GeV. For instance, for $\Lf=10^8\gev$, in order
 to push $m_t$ to $170 \gev$ we should take
$\Ga{33}\sim 2.2$. Such a value, as discussed above, would lead
to a large $\Ga{21}>4$
and the loss of perturbative control. In fact, at the same value of $\Lf$,
even in the limiting seesaw
case, i.e. $\Ga{33}\to\infty$ in eq.\ (21), the top quark cannot weigh more
than $176$ GeV.

{ii)} {\underbar {A lower bound on $\sqrt{R}$}}. This is a consequence
of the CDF bound $m_t>91\gev$ [18] (see Table 1b). Notice that the lower bound
on $\sqrt{R}=(m_um_c/m_dm_s)$ is given, in practice, by
$R>[m_t/m_b(m_t)]\simgt 30=R_{ \rm min}$. Then, by using the maximal value
$c/s=10.7$ [1], we get
$$u/d>{\sqrt R_{\rm min}}(s/c)>0.5\eqno(22)$$
The determination of $u/d$ in chiral perturbation theory is still
controversial, essentially because the axial $U(1)_\a$ is broken by the
anomaly and instanton effects may mimic
a substantial portion of the $u$ quark mass [19].
The range of $u/d$ allowed in our model
is consistent with the values reported by Leutwyler in Ref. [17]. On the
other hand,
Donoghue and Wyler [20] give the estimate $u/d=0.3\pm 0.1$. If this last
estimate turns out to be the correct one, then our model is ruled out,
unless $c/s \simgt 14$. By taking $m_c(1\gev)\leq 1.4\gev$ [1], this bound
translates into $m_s\simlt 100 \mev$, which seems too small a value for $m_s$.
(Even though it is very difficult to determine the absolute scale of the
strange quark mass [1]).
\vskip 1.0truecm
\noindent {\bf 4. Inverse Hierarchy Ansatz with charged leptons}
\vskip 0.5 truecm
Lepton masses have a {\it mixed} behaviour, $m_\mu/m_e \sim \eps{u}$
and $m_\tau/m_\mu \sim \eps{d}$, which seems hard to explain by means
of only one expansion parameter (see Fig.\ 1).
On the other hand, we saw that the large value of the Cabibbo angle requires
a deviation from the genuine radiative pattern in the quark sector. In fact
the off-diagonal entries in the mass matrix affect sizably the masses of the
first two families.  Here we want to show that the same mechanism allows to
explain charged lepton masses by means of one complex expansion parameter
$\eps{e}$.

We simply postulate the inverse hierarchy form of eq. (15) for all mass
matrices\footnote {$^{8)}$}{We set $C=1$ by rescaling $\ga{23}$. This
does not affect the generality of the ansatz.}
$$ (\Mhat_{k})^{-1}=m^{-1}\Bigl \{\ph{1}+\eps{k}\,\, \gamma^\t\,
\ph{1} \,\gamma+\vert\eps{k}\vert^2\,\, \gat^2\, \ph{1}\,\gamma^2
 \Bigr \}\eqno(23)$$
where $k=u,d,e$. This ansatz is simple: the difference between
$u$-, $d$- and $e$-type fermions is parametrized by the $\eps{k}$ only.
The first family plays the role of a {\it mass unification} point.

The eigenvalues of quarks at the Flavour scale $\Lambda_F$ are the same as
those in eq.\ (17) (with $C=1$) while for leptons we have the analogous
formulae
$$ (m/e)=|1+\eps{e}\ga{11}^2|
\qquad (m/\mu)={{|\eps{e}|\ga{12}^2}\over {|1+\eps{e}\ga{11}^2|}}
\qquad (m/\tau)=|\eps{e}|^2\ga{12}^2\ga{23}^2. \eqno(24)$$
Comparing the above equation with eq.\ (17a), we see that the experimental
relation $\tau\simeq b$ implies $|\eps{d}|\simeq |\eps{e}|$. Then
$|\eps{e}|\ga{11}^2\simeq 1$ as in the down sector, so that $e$
and $\mu$ receive in general sizable corrections from this term. In particular,
when the phase of $\eps{e}$ is close to zero, we have that $\mu$ is enhanced
and $e$ is decreased as required by the experimental data (Fig. 1). Eq.\ (24)
gives a mass formula involving leptons and analogous to eq.\ (20) for quarks.
We can write it in terms of down-type quark and lepton masses
$${\tau \over b}=\left |{\eps{d}\over \eps{e}}\right |^2
=\left ({{e\mu}\over {ds}}\right )^2.\eqno(25)$$
Running down this relation from the Flavour scale $\Lambda_F$,
we get a prediction for $m_sm_d$
$$m_sm_d=m_e m_\mu{\sqrt {m_b\over m_\tau}}{{z_t^{1/4}
\eta_s^2 \eta_b^{-1/2} \eta_\f^{3/2}}}
\simeq {277}\, {z_t^{1/4}\eta_\f^{3/2}}\,(\mev)^2.\eqno(26)$$
Where $\eta_s$ is the QCD running of masses from  $m_t=130\gev$ to 1 GeV, and
$\eta_b$ is the analogous quantity from $m_t$ to $m_b$. The remaining $z_t$
and $\eta_\f$ account for the running from $\Lf$ to $m_t$:
$\eta_\f=(\alpha_s(m_t)/\alpha_s(\Lf))^{4/7}$ represents the effects of pure
QCD. While $z_t$ gives the effects of the top quark Yukawa coupling [15]. For
$m_t<180\gev$ the effects of $z_t$ in eq.\ (26) are always $<5\%$, so that we
neglect it. The quantity $m_sm_d$ for different values of $\Lf$ and the
corresponding results for the light quark masses are displayed in Table
2. We have let $\Lf$ go up to $10^{16}\gev$.\footnote{$^{9)}$}{
Notice that even in a model like the one of the previous section it is
meaningful to run up to such a scale. The fact is that, in comparing quarks to
leptons, QCD running becomes effective above $v_\r$.}
The values reported correspond to  $\alpha_s(M_\z)=0.11$, and the $\eta$'s have
been deduced from Ref. [16].
Notice that the prediction for $m_s$ (and hence $m_d$) is increased
by about $30\%$ when $\alpha_s(M_\z)=0.13$.
 The lower bounds on $m_u/m_d$ correspond to
the limit $R=R_{\rm min}=30$. We also get an upper bound for
$m_u/m_d$. Indeed, from eqs. (17a) and (24) we obtain
$$ |1+\eps{d}\ga{11}^2|=|1+\eps{e}\ga{11}^2|
\left ({{es}\over{\mu d}}
\right )^{1/2}\left ({\tau\over b}\right)^{1/4}
<\left (x+{1\over x}|\eps{d}|\ga{11}^2\right )
\left ({{m_em_s}\over{m_\mu m_d}
}\right )^{1/2} \eqno (27)$$
where $x=(\eta_b \eta_F m_\tau / m_b)^{1/4}=0.9-1.1$, for $\Lf$
varying between $10^4$ GeV and $10^{16}$ GeV. To derive the inequality
in eq. (27) we have used eq.\ (25) for  $|\eps{e}/\eps{d}|$.
Then rather independently on $\Lf$ we obtain
$u/d= |1+\eps{d}\ga{11}^2|/|1+\eps{u}\ga{11}^2|<0.75$,
by considering that $|\eps{d}|\ga{11}^2 \simeq |V_{us}|^2(s/d)\simeq 1$ due
to eq.\ (17b) and $|\eps{u}/\eps{d}|<1/6$ due to eq.\ (19).

 The ansatz (23) gives correct predictions by taking the first family as a
{\it mass unification} point (namely, by taking the same normalizing
factor $1/m$
for $u$, $d$, and $e$ type fermions). We can also reverse the
argument, by testing different factors $1/M_i$ in eq.\ (23), instead
of the same $1/m$. It is then interesting to see what values for
$M_i$'s are implied by our experimental knowledge on fermion masses.
We have $(M_d/M_u)^3=(t/b)(ds/uc)^2$ and  $(M_d/M_e)^3=(\tau/b)(ds/e
\mu)^2$. For instance for $\Lf=10^{12}$, after rescaling masses, we get
$$\eqalign {{M_d\over M_u}&\simeq z_t^{-1/3}\left ({m_t\over {120\gev}}
{{4.25\gev}\over m_b} \right)^{1/3}
\left ({m_s\over {130\mev}} {{1.35 \gev}\over m_c}
\right)^{2/3}\left (0.6{m_d\over m_u}\right)^{2/3}\cr
{M_d\over M_e}&\simeq z_t^{-1/6}
\left ({{4.25 \gev }\over m_b}\right )^{1/3}
\left ({m_s\over {130\mev}}\right)^{4/3}\left (20{m_d\over m_s}\right)^{2/3}
\cr} \eqno(28)$$
where we have put the lepton masses to their values. It is remarkable to see
that, by varying the masses of eq.\ (28) in their allowed experimental ranges,
all  $M_i$'s remain equal within a factor of 2.

\vskip 1.0truecm
\noindent {\bf 5. Discussion}
\vskip 0.5truecm
Motivated by the observation of the {\it inverse hierarchy pattern}
$1/m_i^{u,d}\sim \eps{u,d}^{i-1}/m$ in the quark spectrum and by the fact
that an up-down symmetry could be responsible for the smallness of the
CKM mixing angles, we have constructed a realistic model for quark masses.
The key features of this model are:
\item {i)} A universal seesaw mechanism [8,10]:
the masses of the ordinary quarks $q$ are induced via their mixing with
ultraheavy families of weak isosinglet fermions $Q$.
\item {ii)}  The mass matrices of the $Q$'s are generated by a {\it charge
diagonal} radiative cascade [4,7]: the 1$^{st}$ family gets its mass at tree
level, while the 2$^{nd}$ and 3$^{rd}$
get their masses at 1- and 2-loops, respectively.
\item {iii)} The left-right $P_{\l\r}$ and  up-down $\ud$
\footnote{$^{10)}$}{In fact,
we used a discrete up-down symmetry just for simplicity. $\ud$ can be
easily extended to a local ``isotopic'' invariance via the embedding
$U(1)_L\times U(1)_R\times \ud\subset SU(2)_{L}^Q\times SU(2)_{R}^Q$.
Again, for simplicity we have imposed $CP$-invariance.
Our results are unchanged in the case of complex Yukawa coupling constants. }
discrete symmetries. These are {\it softly}
or {\it spontaneously}  broken in the Higgs potential.

The heaviest $Q$-family $(U,D)$ is unsplit because of the
$\ud$ symmetry. The lighter $Q$'s get their masses from
loop-diagrams involving the breaking of $\ud$ and are
thereby split. We considered a particular, but reasonable range for the
masses of the scalar fields involved in mass generation (see the discussion
below eq. (11)). In this range, the $Q$ mass matrices
are simply determined by two
expansion parameters $\eps{u}$ and $\eps{d}$. These are in general complex
and $\eps{u}\not =\eps{d}$, due to the breaking of $CP$, $P_{\l\r}$ and $\ud$.
The $SU(2)_\r$ breaking scale plays the role of the Flavour
scale. The $Q$ masses are such that $v_\r\sim M_{\u_3}<M_{\d_3}<...<M$.
The ratio $v_\r/M$ is determined by $m_u/v_\l\sim 10^{-5}$.
As a result, the $q$'s mass matrices, given by a seesaw mixing with the
$Q$'s, have the {\it inverse hierarchy} form displayed in eq. (15).

Once again we would like to stress that in our approach the ordinary
light quarks are just spectators of the phenomena that determine the
flavour structure. This structure arises in a sector of heavy fermions
and is transferred to the light ones at their decoupling. This explains
why flavour changing effects are suppressed in low energy phenomena.

The mass matrices of eq.\ (15) reproduce the quark spectrum
(see Fig. 1) and mixing angles in a very economical way. The big
differences between $u$- and $d$-type quark masses  are essentially
determined by one parameter $|\eps{d}/\eps{u}|=\sqrt{R} \sim 6-8$.
Upper bounds for $m_t$ and $m_d/m_u$ are then obtained. Rather independently on
the Flavour scale $\Lf=v_\r$, by invoking reasonable perturbativity
requirements, we have $m_t<150\gev$. On the other hand, without further
assumptions, we obtain $(m_um_c/m_dm_s)>6$. Then the relation $m_u/m_d>0.5$
is obtained by using the strange quark mass range given in Ref.\ [1].
 The experimental input
$m_u/m_d<1$ forces the the Cabibbo angle to be {\it large}: $V_{us}\sim
(m_d/m_s)^{1/2}$ instead of  $V_{us}\sim m_d/m_s$ as one generally expects
within the radiative scenario. Moreover, the correct value
$V_{us}\simeq 0.22$ does not imply the loss of perturbativity,
in contrast to the previous models of radiative mass generation.

The model shows that the inverse hierarchy pattern can be obtained in
an ordinary field theory. Having in mind the idea of some
symmetry between quarks and leptons,\footnote{$^{11)}$}
{The
question of whether the $SU(4)$ quark-lepton symmetry of Pati-Salam [21]
could provide this ansatz requires further speculations.
The $T$-scalars of $SU(4)$ contain leptoquarks, that violate in general
the {\it charge diagonality} of the radiative cascade. Also, $SU(4)$ or
any GUT extension brings neutrinos into the game. The related physical
consequences are under study.} eq.\ (15) can be generalized to charged
leptons, in the inverse hierarchy ansatz of eq.\ (23). In this ansatz the
differences between $u$-, $d$- and $e$-type fermions are simply parametrized
by three complex numbers $\eps{u}$, $\eps{d}$ and $\eps{e}$.
In fact, in eq.\ (23) we have kept the same notation of eq.\ (15) just for
convention: actually $\ph{1}$, $\ph{2}=\gamma^\t \ph{1}\gamma$ and
$\ph{3}=\gat^2 \ph{1} \gamma^2$  parametrize the most general set of three
symmetric and real matrices of rank 1. Then, eq.\ (23) contains 11 relevant
parameters to calculate 13 observables (9 masses + 4 angles).
Hence, we have the two relations in eq.\ (20,25). (Notice, however, that
even by taking $\ph{i}$ hermitean these relations remain; this simply means
that $CP$-invariance is not essential for our results.) The predictions
 can be phrased as $m_s=100-150 \mev$ and $m_u/m_d=0.35-0.75$.

We find it amusing that the radiative idea, implemented with the simplest
symmetries ($P$, $CP$, ``isotopic'' and ``lepton-quark'' symmetries),
can explain the key features of the fermion mass spectrum and weak mixing.
Notice, that we did not exploit any horizontal structure,
 in contrast to all known predictive frameworks for fermion masses
(see e.g. [11,22]). Clearly, a {\it clever} horizontal structure would
only enhance the predictive power of our approach.

Last but not least, we wish to remark that our approach can automatically
solve the strong CP-problem,  without introducing an axion. In fact, due
to $P/CP$-symmetries the coefficient $\Theta_{QCD}$ of $\ggd$ is zero. In
addition, the contribution to $\Theta$, that arises from the determinant of
the whole mass  matrix of the fermions $q,Q$ and $p,n$, is vanishing at tree
level. This is a consequence of $P/CP$ and of the seesaw structure [23].
Higher order corrections to the mass matrix can provide a
$\Theta$-term in the range $10^{-9}-10^{-10}$ which is of interest for the
search of the the neutron electric dipole moment.
\vskip 1.0truecm
{\bf Acknowledgements}

\vskip 0.5truecm
Our work has greatly benefited from discussions with
R. Barbieri, J. Bernabeu, G. Dva\-li, P. Fayet,
H. Fritzsch, L. Hall, H. Leut\-wyler, M. Luty,
A. Masi\-ero, R. Mohapatra, V. Ru\-ba\-kov, R. R\"uckl, S. Po\-kor\-ski,
M. Shif\-man, A. Smilga, U. Sarid,
G. Senjanovi\'c, R. Sundrum,
K. Ter-Mar\-ti\-ro\-syan, V. Zak\-ha\-rov, G. Zoupanos and many others.
The work of Z.G.B. is supported by the Alexander von Humboldt Foundation
and that of R.R. is supported by INFN.

\vskip 1.5truecm
{\bf Appendix }
\vskip 0.5truecm

\noindent {We} report the expression of the function $H(r)$.
Let us define:

$$f(r)=\int_{0}^{\infty} {dx\over x(x+r)}\ln (x+1)=
{1\over r}\left \{\ln r \ln |1-r| +\Re \li_2(r)\right \} \eqno A.1$$
where $\Re \li_2$ is the real part of the Euler dilogarithm. Then
$H$ is given by:
$$H(r)={2\over \ln^2r}\bigl[f(r)+f(1/r)-{\pi^2\over 3}\bigr].\eqno A.2$$
$H$ is a positive function which reaches its maximum at $r=1$:
$H(1)=\pi^2/3-3\simeq 0.290$. In the range $1<r<\infty$, $H$  decreases
monotonically and very slowly to zero  (see Fig. 3).

\vfill
\eject

\def\prl#1#2#3{Phys.Rev.Lett. {\bf #1} (#2) #3}
\def\prd#1#2#3{Phys.Rev.  {\bf #1} (#2) #3}
\def\pl#1#2#3{Phys.Lett. {\bf #1} (#2) #3}
\def\npb#1#2#3{Nucl.Phys. B {\bf #1} (#2) #3}
\def\n#1#2#3{{} {\bf #1} (#2) #3}

\ref

\item{[1]} H.Gasser and H.Leutwyler, Phys. Rep. {\bf 87} (1982) 77.
\item{[2]} S.Weinberg, \prd{D5}{1972}{1962}; \prl{29}{1972}{388};
\item{} H.Georgi and S.L.Glashow, \prd{D6}{1972}{2977}; \n{D7}{1973}{2457};
\item{} R.N.Mohapatra, \prd{D9}{1974}{3461};
\item{} S.M.Barr and A.Zee, \prd{D15}{1977}{2652}; \n{D17}{1978}{1854};
\item{} S.M.Barr, \prd{D21}{1980}{1424}; \n{D24}{1981}{1895};
\n{D31}{1985}{2979};
\item{} R.Barbieri and D.V.Nanopoulos, \pl{91B}{1980}{369}; \n{95B}{1980}{43};
\item{} M.Bowick and P.Ramond, \pl{103B}{1981}{338};
\item{} R.Barbieri, D.V.Nanopoulos and A.Masiero, \pl{104B}{1981}{194};
\item{} R.Barbieri, D.V.Nanopoulos and D.Wyler, \pl{106B}{1981}{303}.
\item{[3]} B.S.Balakrishna, \prl{60}{1988}{1602};  \pl{214B}{1988}{267};
\item{} B.S.Balakrishna, A.L.Kagan and R.N.Mohapatra, \pl{205B}{1988}
{345};
\item{} A.L.Kagan, \prd{D40}{1989}{173}; E.Ma, \prl{63}{1989}{1042};
\item{} K.S.Babu and R.N.Mohapatra, \prl{64}{1990}{2747}.
\item{[4]} R.Rattazzi, Z.Phys. C {\bf 52} (1991) 575.
\item{[5]} P.Sikivie et al., \npb{173}{1980}{189}.
\item{[6]} H.P.Nilles, M.Olechowski and S.Pokorski, \pl{248B}{1990}{378}.
\item{[7]} Z.G.Berezhiani and R.Rattazzi, \pl{279B}{1992}{124}.
\item{[8]} Z.G.Berezhiani, \pl{129B}{1983}{99}.
\item{[9]} M.Gell-Mann, P.Ramond and R.Slansky, in "Supergravity", ed.
D.Freedman et al., North-Holland, Amsterdam, 1979; T.Yanagida,
KEK lectures, 1979 (unpublished);
\item{} R.N.Mohapatra and G.Senjanovi\'c, \prl{44}{1980}{912}.
\item{[10]} Z.G.Berezhiani, Yad.Fiz. {\bf 42} (1985) 1309 [Sov.J.Nucl.Phys.
{\bf 42} (1985) 825];
\item{} D.Chang and R.N.Mohapatra, \prl{58}{1987}{1600};
\item{} A.Davidson and K.S.Wali, \prl{59}{1987}{393};
\item{} S.Rajpoot, \pl{191B}{1987}{122}.
\item{[11]} S.Dimopoulos, \pl{129B}{1983}{417};
\item{} G.B.Gelmini et al., \pl{135B}{1984}{103};
\item{} Z.G.Berezhiani,  \pl{150B}{1985}{177};
\item{[12]} R.N.Mohapatra and J.C.Pati, \prd{D12}{1975}{2558};
\item{} G.Senjanovi\'c and R.N.Mohapatra, \prd{D12}{1975}{1502}.
\item{[13]} S.L.Glashow and S.Weinberg, \prd{D15}{1977}{1958};
\item{} E.A.Paschos, \prd{D15}{1977}{1964}.
\item{[14]} J.L.Rosner in ``B-decays'', ed. by S.Stone, World Scientific,
Singapore, 1991.
\item{[15]} B.Pendleton and G.G.Ross, \pl{98B}{1981}{291};
\item{} C.T.Hill, \prd{D24}{1981}{691}.
\item{[16]} G.W.Anderson et al., OHSTPY-HEP-92-018, September 1992.
\item{[17]} H.Leutwyler, \npb{337}{1990}{108}.
\item{[18]} C.Abe et al., CDF-Collaboration, \prl{68}{1992}{447}.
\item{[19]} D.B.Kaplan and A.V.Manohar, \prl{56}{1986}{2004};
\item{} K.Choi, C.W.Kim and W.K.Sze, \prl{61}{1988}{794};
\item{} K.Choi, \npb{383}{1992}{58}.
\item{[20]} J.F.Donoghue and D.Wyler, \prd{D45}{1992}{892}.
\item{[21]} J.C.Pati and A.Salam, \prd{D10}{1974}{275}.
\item{[22]} H.Fritzsch, \npb{155}{1979}{155}; \pl{184B}{1987}{391};
\item{} H.Georgi and C.Jarlskog, \pl{86B}{1979}{297};
\item{} Z.G.Berezhiani and J.L.Chkareuli, Pis'ma ZhETF {\bf 35} (1982) 494
[JETP Lett. {\bf 35} (1982) 612];
Yad.Fiz. {\bf 37} (1983) 1043 [Sov.J.Nucl.Phys. {\bf 37} (1983) 618];
\item{} B.Stech, \pl{130B}{1983}{189}; G.Ecker, Z.Phys. C {\bf 24} (1984) 353;
\item{} M.Shin, \pl{145B}{1985}{285}; J.L.Chkareuli, \pl{246B}{1990}{498};
\item{} S.M.Barr, \prl{64}{1990}{353};
\item{} S.Dimopoulos, L.J.Hall and S.Raby, \prl{68}{1992}{1984};
\item{} K.S.Babu and Q.Shafi, \pl{294B}{1992}{235};
\item{} Z.G.Berezhiani and L.Lavoura, \prd{D45}{1992}{934}.
\item{[23]} K.S.Babu and R.N.Mohapatra, \prd{D41}{1990}{1286}.
\item{} Z.G.Berezhiani, Mod.Phys.Lett. {\bf A6} (1991) 2437.

\vskip 2.0truecm

\line {}
\noindent {\bf Figure captions}
\vskip 0.5truecm
\item {Fig.\ 1} Logarithmic plot of fermion masses as a function of the
family number.
Points corresponding to fermions with the same electric
charge are joined. Leptons are given by the dashed line.
The value $m_t=130 \gev$ has been assumed.
The masses correspond to a renormalization scale $\mu= 10^6\gev$.

\item {Fig.\ 2} a) tree level diagram giving mass to $U$ via mixing with $p$.
b) one-loop correction to the mass matrix of the $Q$-fermions. c,d,e)
two-loop corrections. The analogues of d) and e), where a $T_\r$
and a $H_\l$ are respectively exchanged, are not displayed.

\item {Fig.\ 3} Plot of $H(r)$.
\bye